\documentclass[a4paper,fleqn,usenatbib]{mnras}

\usepackage{color}
\usepackage{dcolumn}
\usepackage{bm}
\usepackage{graphicx}
\usepackage{aas_macros}
\usepackage{subfig}
\usepackage{amsmath}
\begin{document}
\label{firstpage}
\pagerange{\pageref{firstpage}--\pageref{lastpage}}

%\title[GAMA Mass Functions]{Galaxy And Mass Assembly (GAMA): Stellar mass functions by galaxy type to 10$^{8}$M$_{\odot}$ from the GAMA-II survey}
%\title[GAMA Morphological Mass Functions]{GAMA-II Stellar Mass Functions by Galaxy Type to $\sim$10$^{8}$M$_{\odot}$}
\title[GAMA Stellar Mass Budget]{Galaxy And Mass Assembly (GAMA): the Stellar Mass Budget by Galaxy Type}

\author[Moffett et al.]{Amanda J. Moffett,$^1$$^\dagger$ Stephen A. Ingarfield,$^1$ Simon P.~Driver,$^{1,3}$ Aaron S. G.
\newauthor Robotham,$^1$ Lee S. Kelvin,$^{2,5}$ Rebecca Lange,$^1$ Uro\v{s} Me\v{s}tri\'{c},$^2$  Mehmet Alpaslan,$^4$
\newauthor Ivan K. Baldry,$^5$ Joss Bland-Hawthorn,$^6$ Sarah Brough,$^7$ Michelle E. Cluver,$^8$
\newauthor Luke J. M. Davies,$^1$ Benne W. Holwerda,$^9$ Andrew M. Hopkins,$^{7}$ Prajwal R.
\newauthor Kafle,$^1$ Rebecca Kennedy,$^{10}$ Peder Norberg,$^{11}$ and Edward N. Taylor$^{12}$ \\
$^1$ICRAR, The University of Western Australia, 35 Stirling Highway, Crawley WA 6009, Australia\\
$^2$ Institute f{\"u}r Astro- und Teilchenphysik, Universit{\"a}t Innsbruck, Technikerstra{\ss}e 25, 6020 Innsbruck, Austria\\
$^3$SUPA, School of Physics \& Astronomy, University of St Andrews, North Haugh, St Andrews, KY16 9SS, UK\\
$^4$NASA Ames Research Center, N232, Moffett Field, Mountain View, CA 94035, United States \\
$^5$Astrophysics Research Institute, Liverpool John Moores University, IC2, Liverpool Science Park, 146 Brownlow Hill, Liverpool, L3 5RF \\
$^6$Sydney Institute for Astronomy, School of Physics A28, University of Sydney, NSW 2006, Australia \\
$^{7}$Australian Astronomical Observatory, PO Box 915, North Ryde, NSW 1670, Australia \\
$^8$Department of Physics, University of the Western Cape, Robert Sobukwe Road, Bellville, 7535, South Africa \\
$^9$Leiden Observatory, University of Leiden, Niels Bohrweg 2, 2333 CA Leiden, The Netherlands \\
$^{10}$School of Physics \& Astronomy, The University of Nottingham, University Park, Nottingham, NG7 2RD, UK \\
$^{11}$ICC \& CEA, Department of Physics, Durham University, Durham DH1 3LE, UK \\
$^{12}$School of Physics, the University of Melbourne, VIC 3010, Australia \\
$^\dagger${E-mail: amanda.moffett@uwa.edu.au} }

% These dates will be filled out by the publisher
%\date{Accepted XXX. Received YYY; in original form ZZZ}

% Enter the current year, for the copyright statements etc.
\pubyear{2015}

\maketitle

\begin{abstract}
  We report an expanded sample of visual morphological classifications from the Galaxy and Mass Assembly (GAMA) survey phase two, which now includes 7,556 objects (previously 3,727 in phase one). We define a local ($z <0.06$) sample and classify galaxies into E, S0-Sa, SB0-SBa, Sab-Scd, SBab-SBcd, Sd-Irr, and ``little blue spheroid'' types. Using these updated classifications, we derive stellar mass function fits to individual galaxy populations divided both by morphological class and more general spheroid- or disk-dominated categories with a lower mass limit of log(M$_{*}$/M$_{\odot}$) $= 8$ (one dex below earlier morphological mass function determinations). We find that all individual morphological classes and the combined spheroid-/bulge-dominated classes are well described by single Schechter stellar mass function forms. We find that the total stellar mass densities for individual galaxy populations and for the entire galaxy population are bounded within our stellar mass limits and derive an estimated total stellar mass density of $\rho_{*} = 2.5 \times 10^{8}$ M$_{\odot}$Mpc$^{-3}$h$_{0.7}$, which corresponds to an approximately 4\% fraction of baryons found in stars. The mass contributions to this total stellar mass density by galaxies that are dominated by spheroidal components (E and S0-Sa classes) and by disk components (Sab-Scd and Sd-Irr classes) are approximately 70\% and 30\%, respectively.
\end{abstract}


\setlength{\extrarowheight}{0pt}

\section{Introduction}

The luminosity and stellar mass functions of galaxies are fundamental observational measurements with significant importance for constraining our combined models of cosmology and galaxy formation. Galaxy luminosity functions have a long history of utility and progression in the field (e.g., see reviews of \citealp{Felten77}; \citealp{BST88}; \citealp{Johnston2011} and references therein). Alongside increasingly sophisticated techniques for modeling stellar populations and improvements in estimates of galaxy stellar masses, interest in translating observed luminosity functions into the more concretely physical galaxy mass functions has grown significantly. The fidelity with which galaxy stellar mass functions can be measured has risen steadily in recent years as we push towards larger statistical samples and probe successively lower limits in galaxy mass (e.g., \citealp{Balogh2001}; \citealp{Cole2001}; \citealp{Bell2003}; \citealp*{BGD08}; \citealp{Baldry2012}). The full galaxy stellar mass function is now commonly found to be well described by a two-component \citet{Schec76} function form (e.g., \citealp{BGD08}; \citealp{Peng2010}; \citealp{Baldry2012}).

The hunt for ever larger statistical samples of galaxies has also proven fruitful for understanding both the galaxy populations and physical processes that drive the observed two-component nature of the galaxy stellar mass function. This two-component structure has now been attributed to separate ``red'' and ``blue'' galaxy populations, each with their own characteristic mass functions, by a number of authors (e.g., \citealp{Peng2010}; \citealp{Baldry2012}; \citealp{Taylor2015}). If such red and blue populations are considered to correspond to ``passive'' and ``star-forming'' classes, then as suggested by \citet{Peng2010}, a model of galaxy quenching depending on both environment and galaxy mass seems to provide a plausible physical description of the observed galaxy mass function form.

A variety of other population-based divisions of the galaxy stellar mass function have provided further insights into the galaxy demographics that underpin its form, including divisions by galaxy morphology/structure (e.g., \citealp{Bernardi2010}; \citealp{Bundy2010}; \citealp{Vulcani2011}; \citealp{Kelvin_mfunc}) and by the differing environments that galaxies inhabit (e.g., \citealp{Balogh2001}; \citealp{Bolzonella2010}; \citealp{Calvi2013}). Galaxy morphology and structure are a topic of particular interest, as they are thought to be intimately tied to a galaxy's formation history, with spheroidal structures largely arising from dissipationless processes such as dry mergers (e.g., \citealp{Cole2000}) and disk-like structures largely arising from dissipational gas physics processes (e.g., \citealp{FE1980}). The correlation between galaxy morphology and other demographic properties such as environment is also well known (e.g., \citealp{Dressler80}). Significant strides have been made in quantifying the evolution of the galaxy stellar mass function with redshift as well (e.g., \citealp{Pozzetti07}; \citealp{Elsner08}; \citealp{Marchesini09}; \citealp{Ilbert2013}; \citealp{Grazian2015}), but observational limits make the evolution in the detailed form of the stellar mass function over cosmological time still very challenging to constrain.

In this work, we focus on the low-redshift galaxy stellar mass budget in the Galaxy and Mass Assembly survey \citep{GAMAsurv,Driver2011}. We extend the work of \citet{Kelvin_mfunc} on separating galaxy stellar mass functions by morphology to both a fainter magnitude limit ($r < 19.8$ compared to $r < 19.4$ mag) and a larger sky area (180 deg$^2$ compared to 144 deg$^2$). {\color{black}{This expansion critically allows us to approximately double the sample size of \citet{Kelvin_mfunc}, and we further incorporate a sliding mass-limited sample selection, allowing us to probe a mass limit one dex lower than the earlier work. We take advantage of the expansion in sample size and mass range}} to explore the question of whether or not our stellar mass census is bounded for the total and for individual galaxy types, finding that total stellar mass density is generally well constrained by our sample. We first summarize our data and analysis methods in \S \ref{samp} and \S \ref{meth}. We then present our separate morphological-type stellar mass function fits and derive estimates of the total galaxy stellar mass density and mass breakdowns by morphological type in \S \ref{res}. We briefly summarize and discuss these results further in \S \ref{conc}.

A standard cosmology of ($H_0$, $\Omega_m$, $\Omega_{\Lambda}$) = ($70$ km s$^{-1}$ Mpc$^{-1}$, 0.3, 0.7) is assumed throughout this paper, {\color{black}{and h$_{0.7}$$=$$H_0$/($70$km s$^{-1}$ Mpc$^{-1}$) is used to indicate the $H_0$ dependence in key derived parameters.}}

\begin{figure}
  \includegraphics[width=0.48\textwidth]{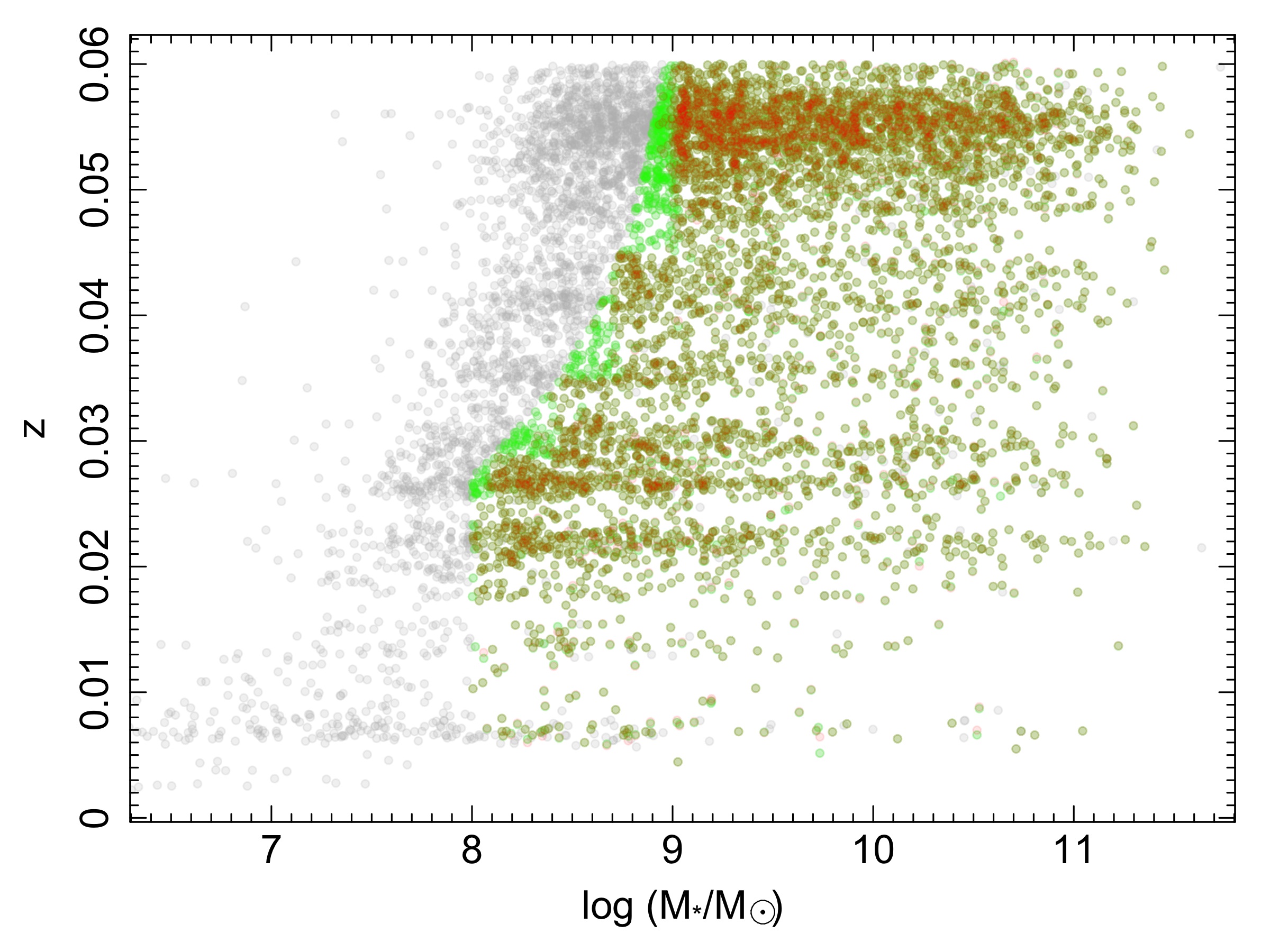}
\caption{The GAMA-II visual morphology sample in redshift versus stellar mass space, where grey points indicate the full classified sample distribution, red points indicate the staggered volume-limited sample of galaxies defined by \citet{Lange2015}, and green points indicate the smoothly defined volume-limited sample we analyse further and describe in \S \ref{MLfits}. {\color{black}{All points are plotted using transparency, and the overlap of red and green symbols, creating darker green, indicates galaxies in both the staggered and smoothly defined volume-limited samples.}}}
\label{fig:samp}
\end{figure}

\section{The GAMA-II data}
\label{samp}
Our data is taken from the Galaxy and Mass Assembly survey phase two, known as GAMA-II.  GAMA is a combined spectroscopic and
multi-wavelength imaging survey designed to study spatial structure in
the nearby ($z < 0.25$) Universe on kpc to Mpc scales (see Driver et
al. 2009, 2011 for an overview and \citet{Hopkins_spec} for details of the spectroscopic data). The survey, after completion of
phase 2 \citep{GAMADR2}, consists of three equatorial regions (and two other
regions) each spanning approximately 5 deg in Dec and 12 deg in RA, centered in RA at approximately
9$^h$ (G09), 12$^h$ (G12) and 14.5$^h$ (G15).  The three equatorial regions, amounting to a total sky area of
180 deg$^2$, were selected for this study as they represented a simple
expansion of the GAMA-I regions used in previous work (Kelvin et
al.~2014a,b) in terms of area (180 deg$^2$ compared to 144 deg$^2$).
The GAMA spectroscopic redshift survey is $>98$\% complete to $r < 19.8$ mag in all
three equatorial regions \citep{GAMADR2}.

% include grid of example classification images
\begin{figure*}
\begin{tabular}{cccc}
\subfloat[star]{\includegraphics[width = 1.5in]{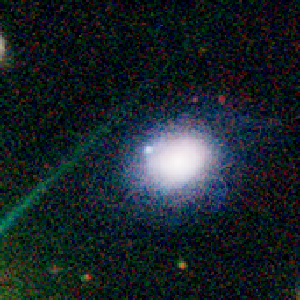}} &
\subfloat[star]{\includegraphics[width = 1.5in]{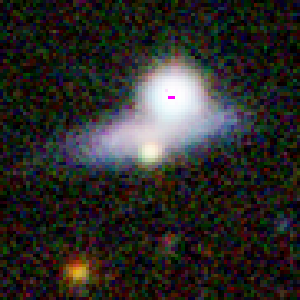}} &
\subfloat[LBS]{\includegraphics[width = 1.5in]{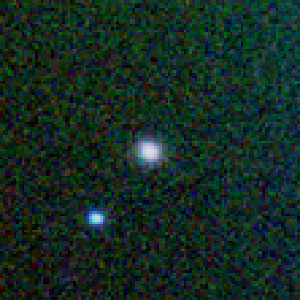}} &
\subfloat[LBS]{\includegraphics[width = 1.5in]{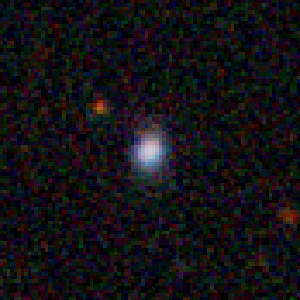}} \\
\subfloat[E]{\includegraphics[width = 1.5in]{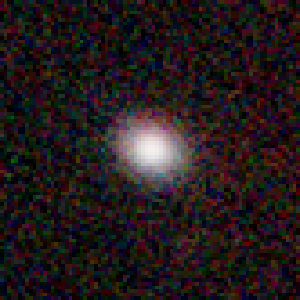}} &
\subfloat[E]{\includegraphics[width = 1.5in]{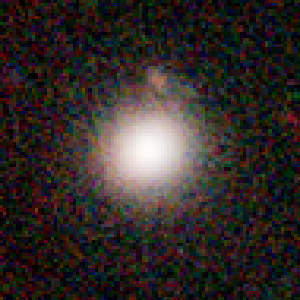}} &
\subfloat[E]{\includegraphics[width = 1.5in]{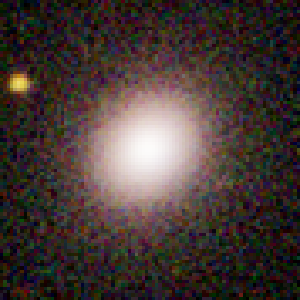}} &
\subfloat[E]{\includegraphics[width = 1.5in]{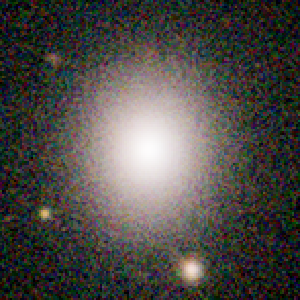}} \\
\subfloat[S0-Sa]{\includegraphics[width = 1.5in]{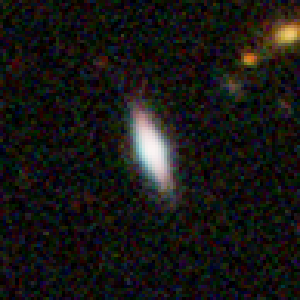}} & %S0-Sa
\subfloat[S0-Sa]{\includegraphics[width = 1.5in]{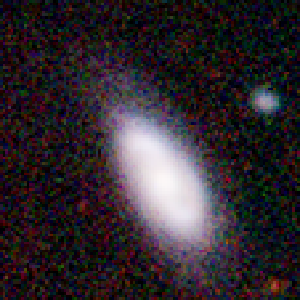}} &
\subfloat[SB0-SBa]{\includegraphics[width = 1.5in]{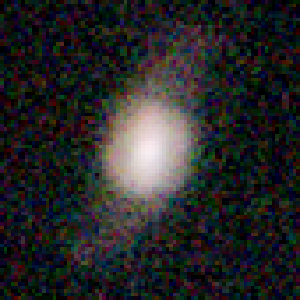}} & % barred
\subfloat[SB0-SBa]{\includegraphics[width = 1.5in]{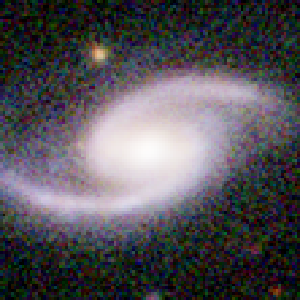}} \\
\subfloat[Sab-Scd]{\includegraphics[width = 1.5in]{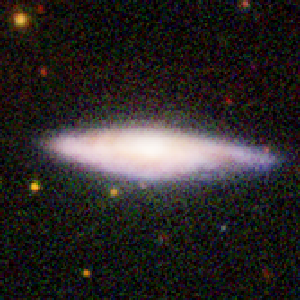}} & %Sab-Scd
\subfloat[Sab-Scd]{\includegraphics[width = 1.5in]{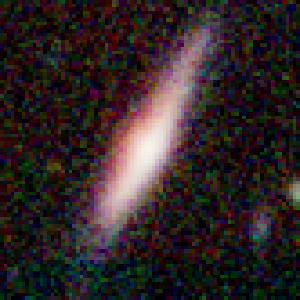}} &
\subfloat[SBab-SBcd]{\includegraphics[width = 1.5in]{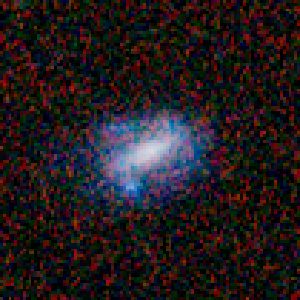}} & % barred
\subfloat[SBab-SBcd]{\includegraphics[width = 1.5in]{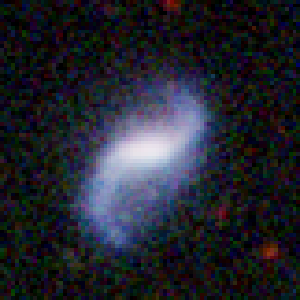}} \\
\subfloat[Sd-Irr]{\includegraphics[width = 1.5in]{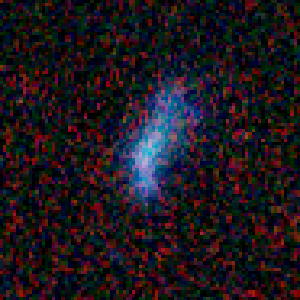}} & %Sd-Irr
\subfloat[Sd-Irr]{\includegraphics[width = 1.5in]{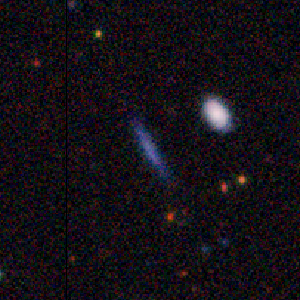}} &
\subfloat[Sd-Irr]{\includegraphics[width = 1.5in]{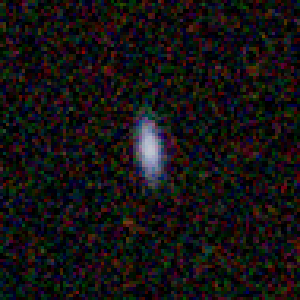}} &
\subfloat[Sd-Irr]{\includegraphics[width = 1.5in]{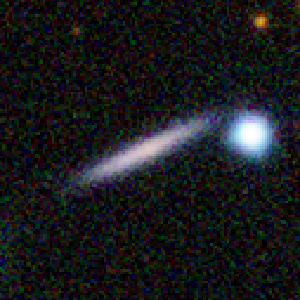}} 
\end{tabular}
\caption{Example three-colour ($giH$) classification images covering a physical distance of 30kpc on a side, as described in \S \ref{class}. Each image is labelled with its assigned type according to the hierarchy illustrated in Fig. \ref{fig:fig2}.}
\label{fig:classims}
\end{figure*}

\begin{figure*}
  \includegraphics[width=0.9\textwidth]{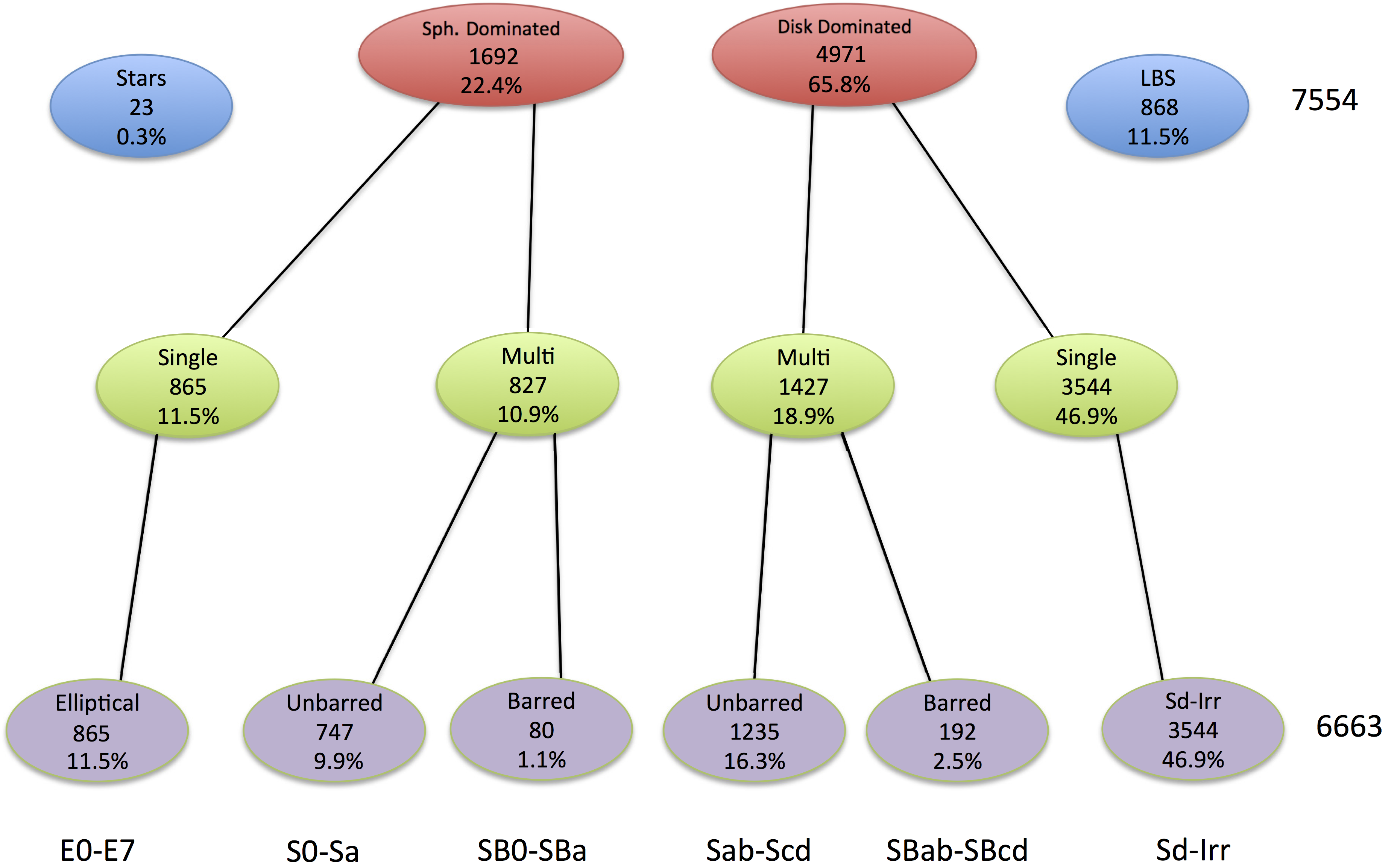}
\caption{Morphological classification hierarchy used to filter the GAMA-II sample of 7,554 galaxies into their
appropriate class. The label LBS indicates the ``little blue spheroid'' galaxy class. Beneath each label are the number of galaxies in the master classification bin for that group and an indication of the fraction of our total sample this group constitutes (fractional contributions as a function of stellar mass are illustrated along with associated error bars in Fig. \ref{fig:nfrac}).  Objects classified as artefacts (2) are not shown.}
\label{fig:fig2}
\end{figure*}

Using the latest version of the GAMA-II tiling catalogue ({\color{black}{TilingCatv44}}; \citealp{Baldry_tiling}) we select from the three equatorial GAMA-II regions a sample of 7,556 {\color{black}{galaxy-like targets (survey\_class $\geq$ 1)}} whose local flow-corrected redshifts lie in the range $0.002 < z < 0.06$ with an associated normalized redshift quality nQ$>2$ (i.e., good for science). The upper redshift limit is motivated by our ability to resolve typical galaxy bulges in Sloan Digital Sky Survey (SDSS; \citealp{SDSScat}) imaging as discussed in \citet{Kelvin_class}, and the lower redshift limit {\color{black}{is motivated by the exclusion of stars and large relative errors on distance.}} The sample is also defined by extinction corrected SDSS $r$ band Petrosian magnitude of $r < 19.8$ mag. This sample is represented by the grey points in Fig. \ref{fig:samp}). The classified sample is defined with no absolute magnitude or mass cuts, so non-galaxy objects blended with a nearby galaxy can still enter the visual classification sample. Flow-corrected redshifts and redshift qualities are taken from the latest GAMA DistancesFrames catalogue (DistancesFramesv12; \citealp{Baldry2012}).

\section{Methods}
\label{meth}
In this section, we describe the GAMA-II morphological classification procedure and our methods for deriving stellar mass function fits to this sample.

\subsection{GAMA-II morphological classifications}
\label{class}
The entire sample of 7,556 galaxy-like objects is visually classified into Hubble types: elliptical (E),
lenticular/early-type spiral (S0-Sa, barred and unbarred),
intermediate/late-type spiral (Sab-Scd, barred and unbarred),
disk-dominated spiral or irregular (Sd-Irr), and little blue
spheroid (LBS) following the same classification tree pattern as \citet{Kelvin_class}. Non-galaxy objects entering the sample were also visually identified and classified as either ``star'' or ``artefact.'' {\color{black}{Note that the star category does not in general correspond to isolated nearby stars, as our sample objects are required to have extragalactic redshifts, but rather consists of a small composite population of star-galaxy blends and possible star clusters or compact satellites that have appear as essentially point sources in our classification images. Artefacts typically correspond to shredded subunits of a larger galaxy already in our catalog.}}\footnote{{\color{black}{We note that our initial morphology catalog was based on 7941 sources (TilingCatv43 without the survey\_class $\geq$ 1 requirement), some of which were artefacts flagged in prior source catalog classifications with vis\_class=3 \citep{Baldry_tiling}. Shredded subunits identified by the morphology team were added to the vis\_class flag in TilingCatv44 by IKB. Several sources were previously classified as artefacts by the morphology team in cases where the SDSS photometric object was located off centre from the main galaxy body. In 24 cases, sources were reassigned galaxy morphological types in the current catalog by LSK. The current selection requiring survey\_class$\geq$1 excludes nearly all artefacts present in the source catalog.}}}

Galaxies were classified into their appropriate morphological types by visual inspection of three-colour images, created for each source
in our sample of 7,556 objects. The classification images consist of a red colour channel from the VISTA Kilo-degree Infrared Galaxy survey (VIKING; \citealp{VIKING}) $H$\footnote{Note that this represents a departure from GAMA-I classifications which used lower-quality $H$ band data from the UKIRT Infrared Deep Sky Survey
(UKIDSS) Large Area Survey (UKIDSS-LAS; \citealp{UKIDSS}).} and green and blue colour channels from SDSS $i$ and $g$ bands \citep{SDSScat}, respectively. Classification images are scaled using the tanh function and depict a fixed physical size equivalent to 30kpc $\times$ 30kpc evaluated at each galaxy's distance, providing some context for each galaxy's local surroundings. This distance-dependent cutout size represents a departure from \citet{Kelvin_class}, where two classification images with differing angular sizes defined \emph{on-sky} were examined for each galaxy. The chosen image size in this work is equivalent to $\sim26''$ at our maximum distance, which is intermediate between the two fixed sizes used in \citet{Kelvin_class}. Example classification images for various classes are shown in Fig.\ \ref{fig:classims}.

\begin{figure*}
  \includegraphics[width=0.97\textwidth]{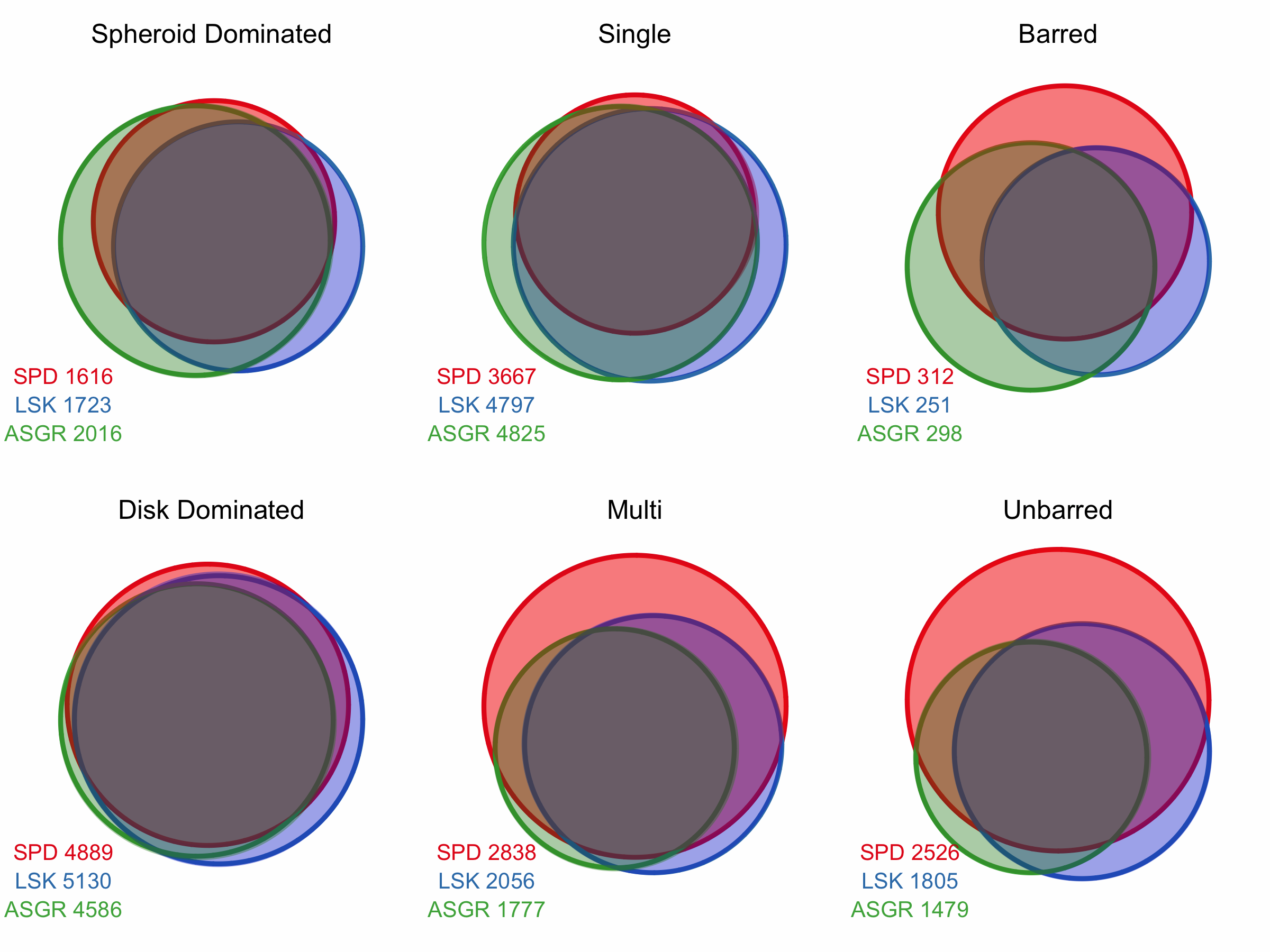}
\caption{Euler diagrams representing the typically high level of agreement between
  the three final visual classifiers (SPD, LSK, and ASGR) for the six main decision tree
  classifications; spheroid dominated/disk dominated, single/multi component and
  barred/unbarred.  The number of objects selected by each classifier
  is shown for each classification category, and the area of each circle is proportional to this classified number.}
\label{fig:fig3}
\end{figure*}

Classification is carried out by placing each galaxy within a specific structural hierarchy. A schematic representation of
this hierarchy is shown in Fig.~\ref{fig:fig2}, with final number counts of each category inset.  Visual classification decisions were initially
made at the top level into five categories: stars, little blue
spheroids (LBS), spheroid dominated, disk dominated, and artefacts.
Objects initially classified as artefacts (2 objects) are not shown
in Fig.~\ref{fig:fig2}, and are not included in any subsequent
analyses (except when looking at classification consistency between
observers in the following section).  At the next level down, both
spheroid dominated and disk dominated objects were further classified
into single-component and multi-component objects.  At the lowest level, multi-component objects
(both spheroid dominated and disk dominated) were further sub-divided
into barred and unbarred categories.  In this manner, all galaxies
were allocated to an appropriate morphological class, from E to
Sd-Irr, as indicated.  The decision tree is essentially binary at each
level (with the exception of the stars and LBS classes). These three levels are spheroid dominated/disk dominated,
single-/multi-component and barred/unbarred.

Three initial classifiers: SAI, UM and RL, independently classified the entire
sample of 7,556 objects, and each set of classifications were subsequently reviewed (or modified where
necessary) by a paired teammate from the original classification team of \citet{Kelvin_class}: SPD, LSK and ASGR.

\subsubsection{Classification Consistency}
A final master classification is assigned based on majority agreement,
in much the same manner as described in \citet{Kelvin_class}.  The
master classification for each object was decided based on a combination of the visual classifications by the three classification teams (SPD/SAI, LSK/UM and ASGR/RL).  Where there was either a two-way or a three-way agreement then the classification
supported by the majority of the classifiers would apply. Only where there was a three-way
disagreement would the master classification default to the classifier
deemed to have the most classification experience (SPD). At the top classification level, there are 194 such three-way disagreements (2.6\%). {\color{black}{A total of 109 objects (1.4\%) were classified as either star or artefact by at least one observer.}}

A visual representation of the level of agreement between classifiers
on the three standard questions (spheroid dominated or disk dominated,
single component or multi component, and barred or unbarred) is shown in
Fig.~\ref{fig:fig3}.  These are area proportional Euler diagrams
which depict both the numbers selected for each classification by the classifier (by the area of each circle) and the level of agreement
between classifiers for each classification (by the areas of overlap
between circles).

Relative to the total galaxy-type sample of 7554, we find the following population fractions are classified with 3-way agreement: spheroid dominated 17.6\% (1331), disk dominated 55.3\% (4174), single 43.3\% (3272), multi 20.0\% (1507), barred 2.2\% (166),
unbarred 14.9\% (1131), Stars 0.2\% (19), LBS 6.1\% (462).  These results
are similar to those reported in \citet{Kelvin_class}, but the generally
lower levels of agreement reflect a higher level of uncertainty in
classifying the larger number of faint objects in the current sample.

The 3-way agreement numbers can be compared with the number of
galaxies within each group in the master classification
(Fig.~\ref{fig:fig2}) to arrive at an overall measure of the degree of
agreement between classifiers in each category: spheroid dominated
78.7\%, disk dominated 83.9\%, single component 74.2\%, multi component 66.9\%, barred
61.0\%, unbarred 56.6\%, stars 60.9\%, LBSs 53.1\%.  Generally there is
good agreement between observers, with the highest consensus in
distinguishing between spheroid-dominated and disk-dominated galaxies.
However, the level of consensus is lowest for distinguishing between barred and
unbarred galaxies, which likely explains the relatively low bar fraction we find ($\sim$12\% for our multi-component galaxies compared to the nearly 30\% bar fraction in disk galaxies found by Galaxy Zoo 2; \citealp{Masters2011}).

From our final combined classifications, just under two-thirds of our sample of 7,554 objects (excluding
artefacts), or 65.8\% (4,971), is visually classified as Disk
Dominated, with spheroid dominated accounting for 22.4\% (1,692).
Additionally, 0.3\% (23) of our sample data are classified as ``stars'',
and 11.5\% (868) as ``little blue spheroids'' (see summary of category counts in Fig.~\ref{fig:fig2}).
Close to half of our sample, 46.9\% (3,554), is visually classified as
Sd-Irr type, with elliptical galaxies accounting for 11.5\% (865) of
the sample.  Spheroid-dominated multi-component systems account for
10.9\% (827) of the sample, of which 9.6\% (80) are visually barred.
Disk-dominated multi-component systems account for 18.9\% (1,427) of
the sample, of which 13.5\% (192) are visually barred.  These
classifications will be used throughout the remainder of this work.

\begin{figure}
\includegraphics[width=0.48\textwidth]{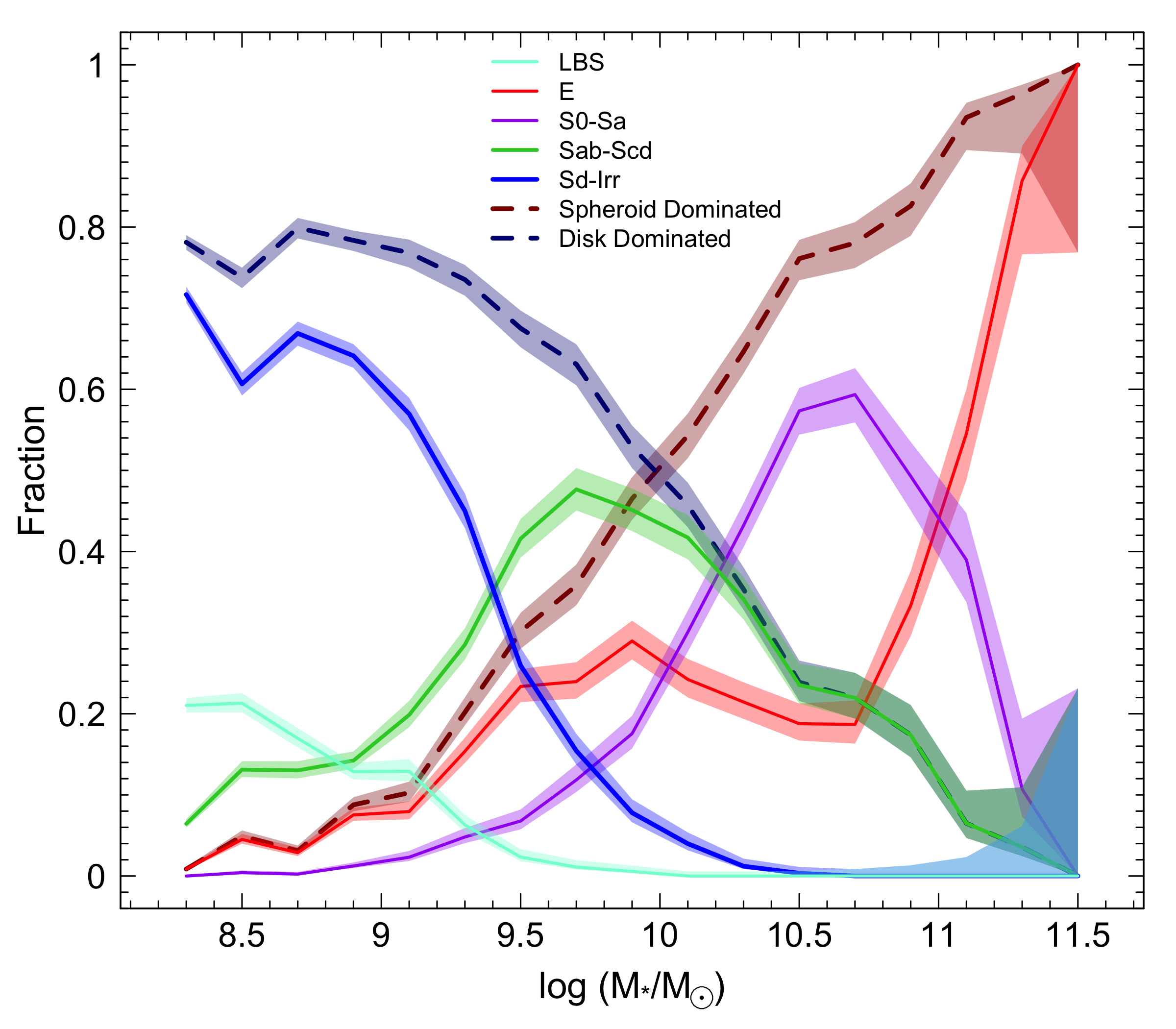}
\caption{\label{fig:nfrac}Fractional contribution of various morphological types to the total by number as a function of stellar mass, with $V/V_{\rm max}$ weights applied. Spheroid-dominated and disk-dominated classes are a combination of multiple morphological classes as indicated in Fig. \ref{fig:fig2}, and as such are complements of one another until the low stellar mass range where LBSs, which we do not place in either category here, become common. Error bars are derived using the \citet{Cameron2011} beta distribution method for estimating binomial confidence intervals.}
\end{figure}

\subsection{Maximum Likelihood Stellar Mass Function Fits}
\label{MLfits}
We use the galaxy stellar mass estimates of \citet{Taylor2011} derived using GAMA optical photometry and stellar population synthesis modeling with a \citet{Chabrier} initial mass function. {\color{black}{To derive total stellar mass estimates, we include the additional mass scaling factors discussed by \citet{Taylor2011} that account for light missed in finite-size GAMA apertures by comparison to S\'ersic measures of total flux from \citet{SIGMA}. We note that the typical mass correction required for our sample galaxies is small (0.007 dex) and that our major results do not significantly change even if the correction factors are omitted.}}

Since our full classified sample is apparent magnitude limited, it has a varying mass limit as a function of redshift (see Fig.\ \ref{fig:samp}). {\color{black}{To maximize number statistics and allow us to extend stellar mass function fits to a lower stellar mass limit than for the single mass-limited sample of \citet{Kelvin_mfunc}}}, we take a similar approach to that of \citet{Lange2015} and define a volume-limited subsample of our data with mass limits that are a sliding function of redshift. \citet{Lange2015} define the appropriate mass limits as a function of redshift to create individual volume-limited sample of GAMA-II that is at least 97.7\% complete and unbiased with respect to galaxy colour, and they make a selection in both redshift and mass intervals to define a series of stepped volume-limited samples (see red points in Fig.\ \ref{fig:samp}). We take this approach one step further and fit a smooth function to the same mass limits as a function of redshift, given by M$_{\rm lim} =$ 4.45 $+$ 207.2$z$ $-$ 3339$z^{2}$ $+$ 18981$z^{3}$, and require the sample we analyse in subsequent sections to have stellar mass greater than the appropriate mass limit evaluated at its redshift (see green points in Fig.\ \ref{fig:samp}).

Due to {\color{black}{a known issue with the automated GAMA photometry that may lead to erroneously bright magnitudes and high mass estimates when a galaxy's aperture is affected by the presence of a nearby bright star}}, we choose to visually inspect the apertures for the 20 most massive galaxies in each morphology category in our sample. For 26 objects, we identified apertures that were unreasonably large given the galaxy sizes, which accordingly led to over-estimated stellar masses for these objects. These objects were primarily in the Sd-Irr class (15 out of the 20 inspected), which is also the most numerous class in our sample. We omit these objects from the sample when performing further stellar mass analysis, {\color{black}{which should cause minimal mass incompleteness due to their small fractional contribution to each class.}}

To derive fits to the stellar mass distributions of various morphologically defined subsamples of GAMA-II, we employ a parametric maximum likelihood fitting method similar to one often used in determining galaxy luminosity functions (e.g., \citealp{SandageML}; \citealp{EfML}). This method has the advantage of fitting galaxy stellar mass distributions without the need to bin the source data. However, the requirement of a parametric function can be disadvantageous if it provides a poor description of the data. Specifically we take an approach similar to that described by \citet{Robotham2010}, where the probability density function (PDF) for each galaxy in mass space is represented by a single \citet{Schec76} type functional form:

\begin{multline} 
  \Phi(\log M)d \log M = ln(10)\times\phi^{*}10^{log(M/M^{*})(\alpha+1)}\\
  \times\exp(-10^{ \log (M/M^{*})})d \log M 
\end{multline}

where M$^{*}$ is the characteristic mass corresponding to the position of
the ``knee'' in the mass function, while $\alpha$ and $\phi^{*}$ refer to the slope of
the mass function at the low mass end and the normalization constant,
respectively. 

\begin{figure*}
  \includegraphics[width=0.8\textwidth]{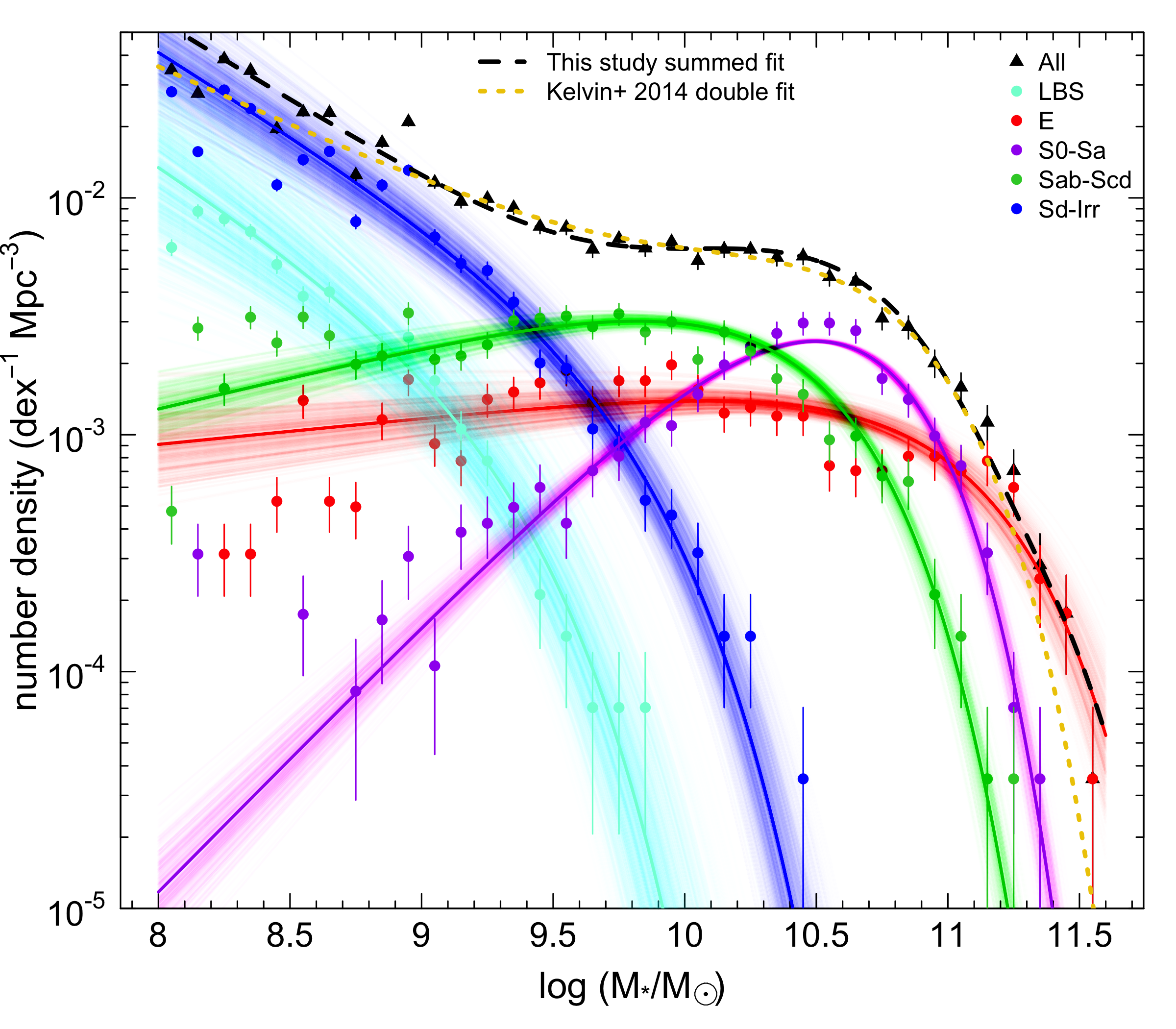}
\caption{Morphological-type stellar mass functions as fit by single Schechter functions, with the sum of these individual fits compared to the total stellar mass function of \citet{Kelvin_mfunc}, which is also very similar in shape to that of \citet{Baldry2012}. Although we do not fit directly to the binned galaxy counts, we show these data counts alongside the fits for illustrative purposes, {\color{black}{using common $1/V_{\rm max}$ weights for objects in 0.3 dex stellar mass bins as defined by}} \citet{Lange2015} and with Poisson error bars on the data counts. Error ranges for the individual MSMF fits are indicated by sampling 1000 times from the full posterior probability distribution of the fit parameters and plotting the resulting sampled mass functions with transparency such that darker regions indicate roughly one sigma uncertainties on the fits. \label{fig:mfunc}}
\end{figure*}

To construct an appropriate likelihood function in this fitting procedure, the PDF that represents each galaxy must integrate to a total probability of one over the stellar mass range. Since our sample is apparent magnitude limited, our stellar mass limit also varies as a function of redshift. As a consequence, each galaxy in our sample has a corresponding mass range over which its PDF is defined, with the lower limit set to the sample mass limit at its redshift. In this way we normalize each galaxy's PDF to account for our redshift-dependent selection function, which makes this approach analogous to the application of $V/V_{\rm max}$ sample weights\footnote{Note that $V_{\rm max}$ values are often used in a density-corrected form to account for variations in large-scale structure along the line of sight. However, our fitting method does not require separate density corrections to account for such variations.}. Accordingly, we determine the appropriate lower integration limits as a function of redshift for this procedure using the GAMA II $V/V_{\rm max}$ correction analysis of \citet{Lange2015}, using a fit to the mass limits as a function of redshift as discussed previously for our fitting sample selection.

The galaxy PDFs are summed over the entire chosen sample to give the likelihood function that is then maximized to derive the most likely Schechter $\alpha$ and M$^{*}$ parameters. We use a Markov Chain Monte Carlo (MCMC) procedure for this analysis, implemented in the contributed \textsf{R} package $LaplacesDemon$\footnote{https://github.com/Statisticat/LaplacesDemon}. We choose to use the Componentwise Hit-And-Run Metropolis (CHARM) algorithm in this package and specify only a flat/uniform prior on fit parameters. We perform a minimum of 10,000 iterations for each fit (fits are also carried out 10 times for each class in order to derive jackknife errors on the fit parameters as discussed in \S \ref{errorcomp}) but also check for convergence using the $Consort$ function of $LaplacesDemon$\ and increase iterations performed for some classes where necessary. Since this procedure does not directly fit for the overall $\phi^{*}$ normalization parameter, we derive this value for each population by requiring that the integrated Schechter function match the summed galaxy number distribution over a mass interval in which galaxy populations are well sampled ($9 <$ log(M$_{*}$/M$_{\odot}$) $< 10$ for all types except Es where we sum up to log(M$_{*}$/M$_{\odot}$) $= 11$ for improved statistics).

\section{Results}
\label{res}

Fig.\ \ref{fig:nfrac} illustrates the variation in frequency of the morphological types in our sample as a function of stellar mass, where barred and unbarred members of the same morphology class are considered together. $V/V_{\rm max}$ weights are used as derived by \citet{Lange2015},\footnote{Note that we have also considered the use of density-corrected $V_{\rm max}$ values to weight our binned data points shown in this figure and in subsequent stellar mass function figures, but we find that the application of this correction makes negligible difference within the given errors on each binned data point.} where galaxies in stepped stellar mass bins of 0.3 dex are assigned identical weights. Weights are not required for log(M$_{*}$/M$_{\odot}$) $> 9$ galaxies.

\begin{figure}
  \includegraphics[width=0.48\textwidth]{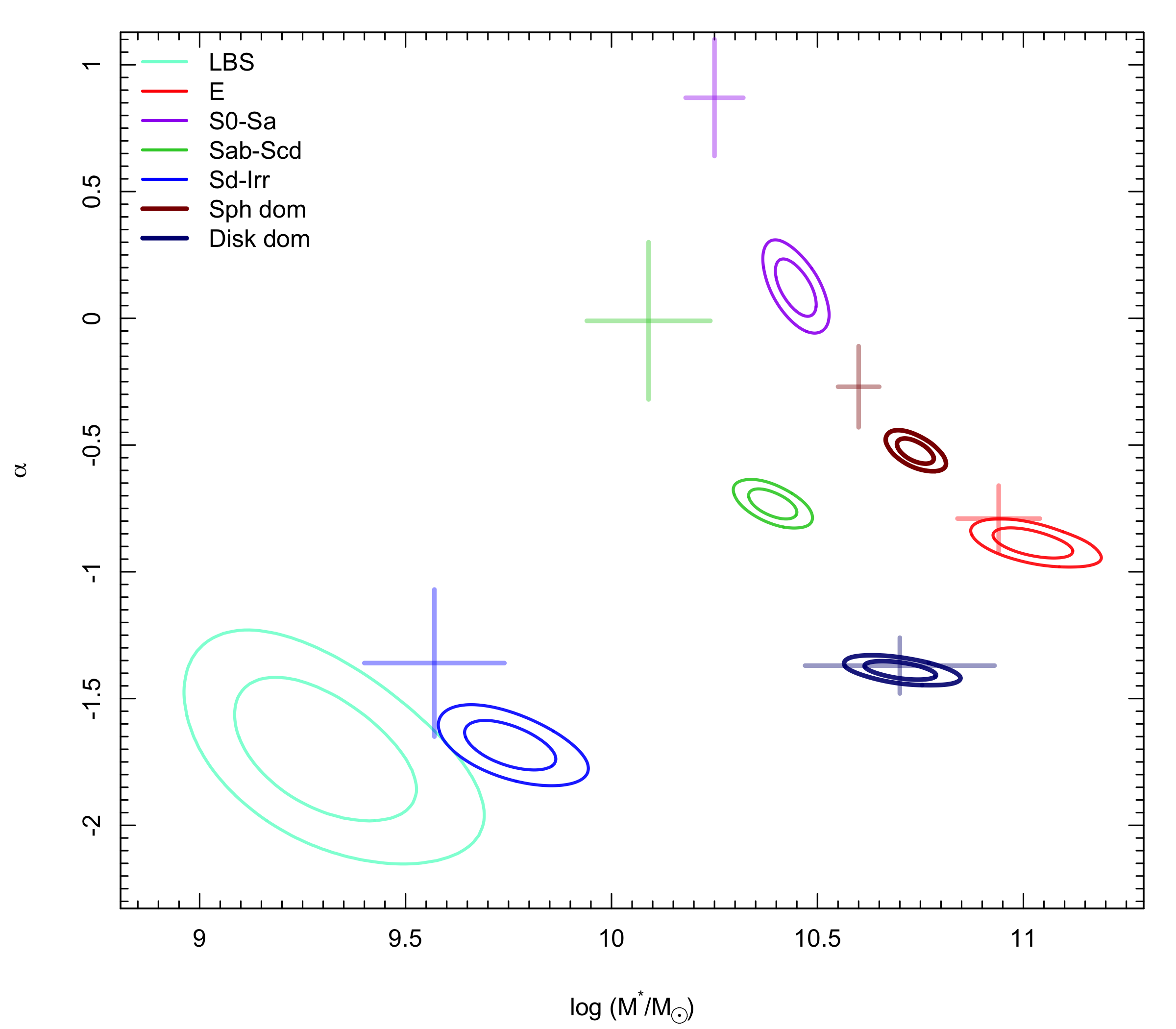}
\caption{Error contours for separate morphological-type and bulge-/disk-dominated stellar mass function fits, shown at the one and two sigma levels, with crosses indicating the values and one-sigma parameter errors derived from jackknife resampling in \citet{Kelvin_mfunc}.  The error contours are derived from a jackknife resampling procedure with 10 subvolumes and consideration of the two-dimensional posterior probability distributions of all resulting fits. \label{fig:conts}}
\end{figure}

As was also seen in the GAMA I analysis of \citet{Kelvin_mfunc}, elliptical galaxies are most common at high stellar masses, while progressively later types reach their peak numerical dominance at lower stellar masses. Similar results are also seen in the all-sky sample of \citet{Conselice2006}. The spiral class of \citet{Conselice2006} also shows similar frequency variation with mass as our Sab-Scd class, peaking in frequency near log(M$_{*}$/M$_{\odot}$) $= 10$, and the \citet{Conselice2006} irregular class likewise behaves similarly to our Sd-Irr class. We find the transition point between numerical dominance of spheroid-dominated and disk-dominated classes occurs around log(M$_{*}$/M$_{\odot}$) $= 10$. Both Sd-Irr and LBS classes exist as primarily ``dwarf'' objects, with near-zero frequency above log(M$_{*}$/M$_{\odot}$) $= 10$ and rising in frequency still at the lowest stellar mass limits of our sample. Such GAMA dwarfs have been studied in greater detail in a number of other works and are largely found to be a star forming population inhabiting primarily low density environments (e.g., \citealp{Brough2011}; \citealp{Bauer2013}; \citealp{Mahajan2015}). 

The LBS class is an interesting case where galaxy color is apparently at odds with typical expectations for roughly early-type morphology, and this class appears to substantially overlap with the blue early-type class of other authors (e.g., \citealp{KGB}; \citealp{Schaw09}). However, at the low stellar mass limits we probe here we find that LBSs occur with substantially higher frequency, reaching $\sim20\%$, compared to the few percent frequencies observed for blue early-types in samples with higher mass limits. \citet{KGB} shows that blue early-type frequency rises strongly with decreasing stellar mass, reinforcing the likely overlap between these two populations. As a result of their typically low stellar masses, the stellar mass distribution of the LBS class, in particular, could not be well constrained using the higher mass limit sample of \citet{Kelvin_mfunc}, but in this work we are able to constrain the stellar mass function for this class.

\subsection{Stellar Mass Functions Divided by Morphology}

Fig.\ \ref{fig:mfunc} shows the derived morphological-type stellar mass function (MSMF) fits for our sample, where we only include galaxies down to log(M$_{*}$/M$_{\odot}$) $= 8$ (fit parameters summarized in Table \ref{tab:tab1}). Below this mass, the GAMA sample is expected to suffer from significant surface-brightness-based incompleteness (see \citealp{Baldry2012} for further details). Single Schechter functions provide good fits to each of our individual morphologically defined populations, and summing our individual MSMF fits provides, as would be expected, an excellent description of the total stellar mass function of this sample (see dashed black line in Fig. \ref{fig:mfunc}). Our total stellar mass function would clearly require \emph{at least} a double Schechter fit to describe well, as has been argued by other authors (e.g., \citealp{BGD08}; \citealp{Peng2010}; \citealp{Baldry2012}; \citealp{Kelvin_mfunc}). In Fig.\ \ref{fig:mfunc}, we show the double Schechter fit of \citet{Kelvin_mfunc} for comparison with our summed fit, and find that they agree reasonably well, although our combined low-mass slope is steeper than that of \citet{Kelvin_mfunc}.

\begin{figure*}
  \includegraphics[width=0.8\textwidth]{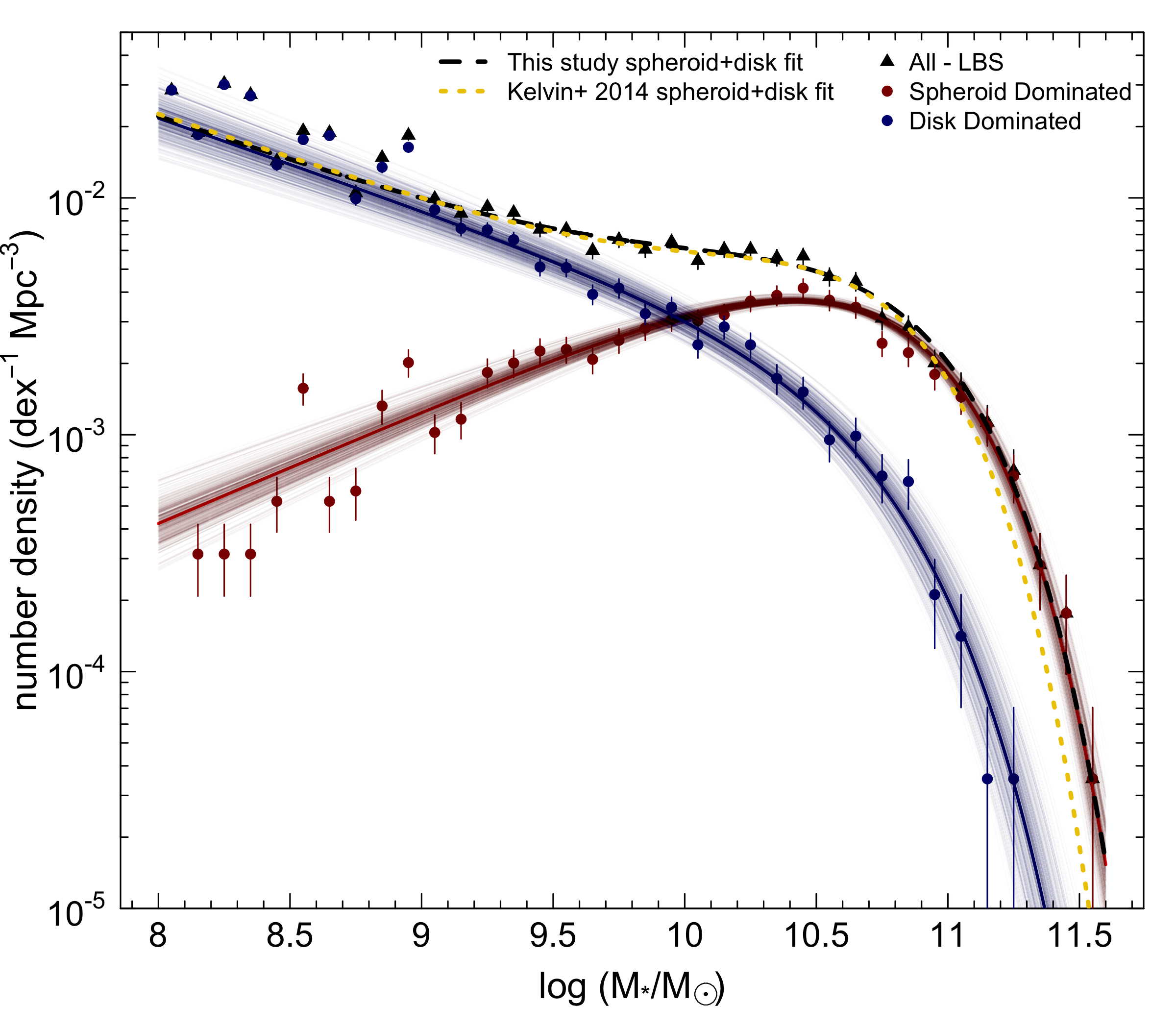}
\caption{Spheroid-dominated (E/S0-Sa) and disk-dominated (Sab-Scd/Sd-Irr) galaxy stellar mass distributions are fit by single Schechter functions (dark red and blue points/lines, respectively), which combined (black dashed line) fit the total stellar mass distribution well. Since LBSs are not included in either of these categories, we show a version of the total stellar mass data points (black triangles) omitting the LBS galaxies for comparison. The combined spheroid- and disk-dominated mass function of \citet{Kelvin_mfunc} is also shown for comparison, which is similar to the combined red and blue galaxy mass functions of \citet{Baldry2012} and \citet{Peng2012}. Data point weights and error ranges are indicated as in Fig.\ \ref{fig:mfunc}. \label{fig:bdfits}}
\end{figure*}

\label{errorcomp}
Compared to \citet{Kelvin_mfunc}, which has a mass limit of log(M$_{*}$/M$_{\odot}$) $= 9$, we find the detailed shapes of our MSMFs differ for some classes. Fig.\ \ref{fig:conts} compares the error contours from our fits to the Schechter fit parameters reported in \citet{Kelvin_mfunc}, where in both cases the error ranges are derived from a jackknife resampling procedure with 10 subvolumes. We consider the full two-dimensional posterior probability distributions of all resulting fits and derive error contours that represent the covariance between parameters. As seen in this figure, the E, Sd-Irr, and {\color{black}{disk-dominated mass}} functions agree well within roughly one-sigma paramter fit uncertainties. For {\color{black}{spheroid-dominated, S0-Sa,}} and Sab-Scd populations, our fit parameters disagree with \citet{Kelvin_mfunc} at greater than two sigma, and the sense of the disagreement is that our M$^{*}$ values are higher and $\alpha$ values are lower than in this earlier work. We test to see if this effect is solely due to the difference in mass limits by performing fits for only log(M$_{*}$/M$_{\odot}$) $> 9$ galaxies in our sample but still find discrepancies with the \citet{Kelvin_mfunc} fits in these cases. Close scrutiny of Fig.\ 3 in \citet{Kelvin_mfunc} reveals that fits with slightly higher  M$^{*}$ (with a correlated change in $\alpha$) could have also been justified by the data in this case, and this difference may be due to the exclusion of low number count bins at high mass from the earlier binned data fitting analysis. Consideration of an additional error contribution from misclassification may also bring these measurements into formal agreement within larger error bars. \citet{Kelvin_mfunc} was not able to constrain the shape of the mass function for the LBS class, and while we are able to do so in the current work, the uncertainties in the fits to this population are still significantly larger than for any other population. It is interesting to note that for both our LBS and Sd-Irr classes, the relatively steep  $\alpha$ values we find approach the level of steepness ($\alpha \sim -1.8$) in luminosity function slope that has been frequently observed in cluster dwarf populations (e.g., \citealp{Driver1994}), despite the fact that our sample is numerically dominated by field galaxies.

\begin{figure*}
  \includegraphics[width=0.8\textwidth]{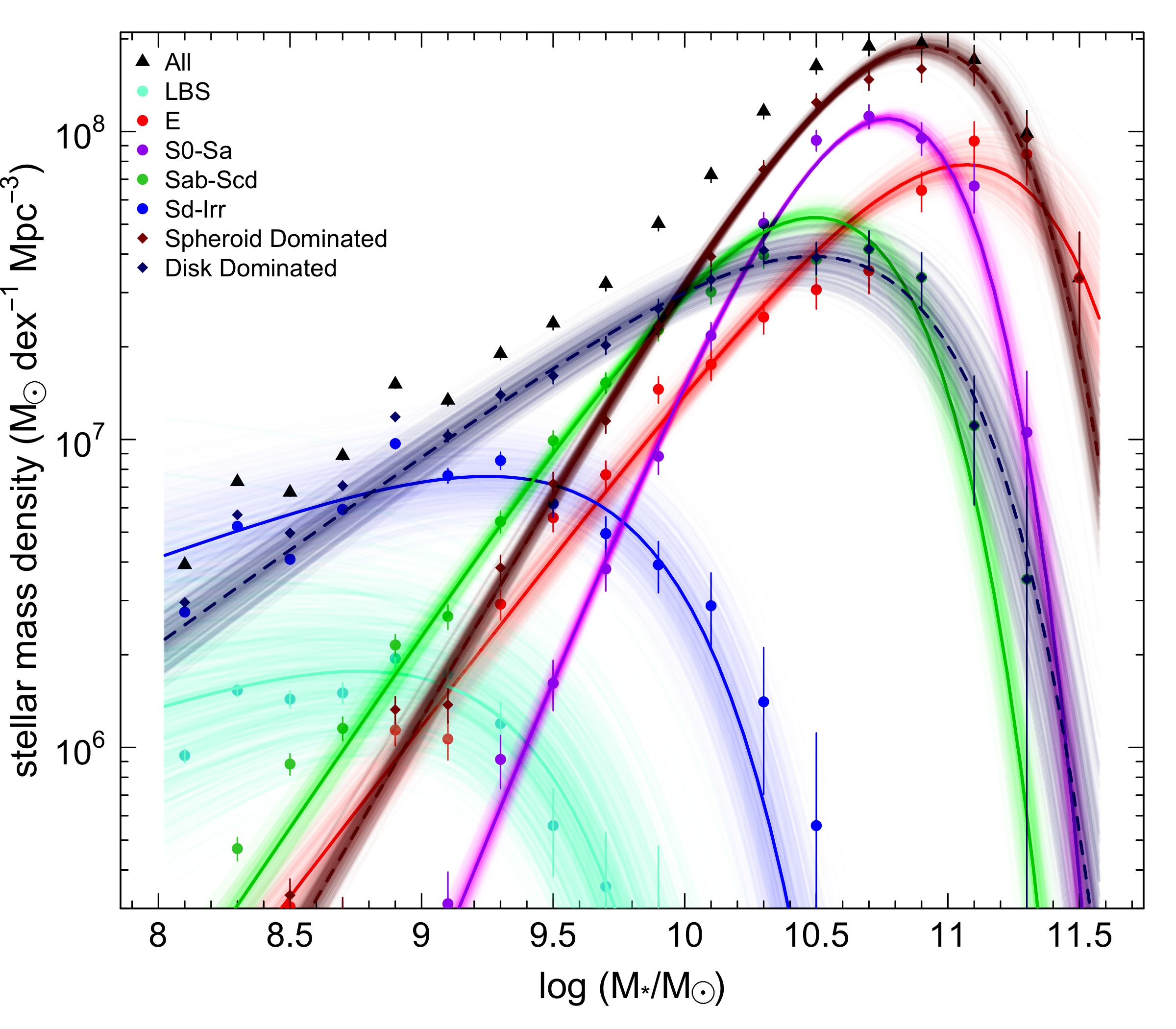}
\caption{Total mass density of our sample and mass density in separate morphological classes, where points indicate the data values (with $1/V_{\rm max}$ weights), and lines indicate values derived from our Schechter function fits. Mass density estimates are bounded for the total and for most individual classes. Error ranges on these fits are indicated as in Figs.\ \ref{fig:mfunc} and \ref{fig:bdfits}.}
\label{fig:totmass}
\end{figure*}

Returning to the presumed double Schechter function nature of the total galaxy stellar mass function, it has been previously observed that individual ``red'' and ``blue'' single Schechter stellar mass functions are similar to the double Schechter function components of the total galaxy stellar mass function (e.g., \citealp{Baldry2012}; \citealp{Peng2012}; \citealp{Taylor2015}). Similarly, we explore whether or not two galaxy structure based categories can effectively describe the total stellar mass distribution of our sample. As in our classification tree scheme (see Fig.\ \ref{fig:fig2}), we define two broad structural classes to be ``spheroid dominated'' (E/S0-Sa) and ``disk dominated'' (Sab-Scd/Sd-Irr) galaxies. Fig.\ \ref{fig:bdfits} shows the results of single Schechter fits to the stellar mass distributions of these two populations. A single Schechter function appears to be a reasonable description of both the spheroid-dominated and disk-dominated populations shown here. If we also include the LBS class in the spheroid-dominated category despite their apparent morphology vs. color mismatch, we obtain an altered spheroid-dominated mass function, with a more two-component appearance and a turn-up to the lowest masses we probe. A similar analysis was performed by \citet{Kelvin_mfunc}, and as can be seen in Fig.\ \ref{fig:conts}, our Schechter fit parameters for the spheroid- and disk-dominated populations are broadly {\color{black}{similar to those}} of \citet{Kelvin_mfunc}. However, in the current study that extends $\sim$1 dex lower in stellar mass than \citet{Kelvin_mfunc}, our spheroid-dominated mass function does have a noticeably less steep decline to low masses than in \citet{Kelvin_mfunc}. \citet{Kelvin_mfunc} found that separate spheroid- and disk-dominated stellar mass functions described the total mass function well and were similar to the red and blue stellar mass functions of \citet{Baldry2012} and \citet{Peng2012}, except for the possible upturn in the red stellar mass function at low masses. {\color{black}{As recently argued by \citet{Taylor2015}, the low-mass upturn observed for red galaxies may not be robust due to applied colour cuts that can group low-mass blue galaxies with the red galaxy population. We find that our combined mass function, minus the LBS population, is well described by the separate spheroid- and disk-dominated mass functions and that only the inclusion of blue in colour LBSs among the spheroid-dominated population would cause an upturn in the mass function of the \emph{spheroid-dominated} population analogous to the \emph{black} population upturn observed by other authors. However, we do not suggest that LBSs are necessarily responsible for the red mass function upturn, as the degree to which such galaxies may have been assigned to the red population in other work has not been investigated in detail here.}}

\subsection{Stellar Mass Densities and Population Fractions}
\label{mbreakdown}
As illustrated in Fig.\ \ref{fig:totmass}, with the low stellar mass limit of a GAMA-II sample we have entered a regime where the total stellar-mass density ($\rho_{*}$) in our volume should be bounded, as both data and stellar mass function fits indicate stellar mass density distributions where the peaks are well sampled. Only in the cases of LBS and Sd-Irr classes does the decline in $\rho_{*}$ appear close to flat, so it is possible that the $\rho_{*}$ value for these classes is not yet as well constrained as for other classes. Table \ref{tab: tab2} summarizes the stellar mass density values we derive from both direct summation of the data with $V/V_{\rm max}$ weights ($\rho_{\Sigma}$) and for integrating our stellar mass function fits ($\rho_{\phi}$). Quoted errors on these values are derived using jackknife resampling and dividing our sample into 10 subvolumes and are defined by $\sigma^{2}=N-1/N \sum_{i=1}^{N} [(x_{i}-x)^{2}]$, with N the number of jackknife subsamples, x the overall best fit parameter, and $x_{i}$ the best fit parameter derived in each subsample. In the case of the direct stellar mass sum estimate $\rho_{\Sigma}$, we also propagate the stellar mass uncertainties derived by \citet{Taylor2011} for each galaxy into our error estimate, but the jackknife error bar is by far the dominant error component. In addition, all such estimates are subject to an additional error term from cosmic variance. Using the methods of \citet{DR2010}\footnote{implemented in an online calculator form at http://cosmocalc.icrar.org/}, we estimate a 22.3\% cosmic variance error for our sample volume. We find that both methods yield total and separate morphological class $\rho_{*}$ estimates that are consistent within the quoted uncertainties, and our $\rho_{*}$ estimates are also consistent with the previously derived values of \citet{Kelvin_mfunc}.

From the integration of our individual MSMF fits, we find a total $\rho_{*} = 2.5 \times 10^{8}$ M$_{\odot}$Mpc$^{-3}$h$_{0.7}$. Taking the same assumptions as \citet{BGD08} for the critical density of the universe and $\Omega_{\rm b}$ (0.045; \citealp{Spergel07}), we find $\Omega_{\rm stars} = 0.0018$ and the fraction of baryons in stars ($\Omega_{\rm stars}/\Omega{\rm b}$) is approximately 4 percent, which is consistent with the estimates in both \citet{BGD08} and \citet{Baldry2012} within quoted errors. Breaking down our total stellar mass density further, we find that approximately 70\% of our total stellar mass is found in spheroid-dominated systems (E and S0-Sa) and approximately 30\% in disk-dominated systems (Sab-Scd and Sd-Irr). However, if we make the simple assumption that our multi-component populations exhibit typical bulge-to-total ratios as calibrated by \citet{GW2008}, we find that total stellar mass would be approximately equally divided between spheroidal and disk-like structures, consistent with prior results (e.g., \citealp{Driver2007LET}). Dividing the stellar mass distribution more finely by morphology, we find that E and S0-Sa classes are the most mass-dominant individual classes, each comprising roughly 35\% of the total stellar mass. The Sab-Scd class is next with 22\% of the total stellar mass, while Sd-Irrs despite their numerical dominance make up only 7.4\% of the total stellar mass. Likewise, our other primarily dwarf class, the LBSs comprise only about 2\% of our total stellar mass.

\begin{table*}
  \caption{\label{tab:tab1} Single Schechter stellar mass function fit parameters for the morphological-type stellar mass functions in Figs.~\ref{fig:mfunc} and \ref{fig:bdfits}. Columns are: the knee in the Schechter function (M$^{*}$), the slope ($\alpha$), and the normalization constant ($\phi^{*}$). Quoted error bars are derived from the spread in each parameter's posterior probability distribution from fits carried out in 10 jackknife resampling iterations.}
\begin{tabular}{cccc} \hline

Population & log(M$^{*}$h$_{0.7}$$^{2}$/M$_{\odot}$) & $\alpha$ & $\phi^{*}/10^-3$  \\
&                  &          & (dex$^{-1}$Mpc$^{-3}$h$_{0.7}$$^{3}$) \\ \hline
E &$11.02 \pm 0.055$ & $-0.888 \pm 0.034$ & $0.865^{+1.0}_{-0.63}$ \\
S0-Sa &$10.44 \pm 0.028$ & $0.127 \pm 0.064$ & $2.90^{+0.15}_{-0.16}$ \\
Sab-Scd &$10.39 \pm 0.034$ & $-0.734 \pm 0.033$ & $2.43^{+0.14}_{-0.14}$ \\
Sd-Irr &$9.755 \pm 0.062$ & $-1.69 \pm 0.054$ & $1.14^{+0.27}_{-0.22}$ \\
LBS &$9.300 \pm 0.12$ & $-1.71 \pm 0.15$ & $0.738^{+0.43}_{-0.28}$ \\
 \hline
Spheroid Dominated &$10.74 \pm 0.026$ & $-0.525 \pm 0.029$ & $3.67^{+0.20}_{-0.20}$ \\
Disk Dominated &$10.70 \pm 0.049$ & $-1.39 \pm 0.021$ & $0.855^{+0.10}_{-0.093}$ \\

\hline
\end{tabular}
\end{table*}

\begin{table*}
 \caption{\label{tab: tab2} Total stellar mass densities and stellar mass densities for each morphological class, derived both by summation of data with $V/V_{\rm max}$ weights ($\rho_{\Sigma}$) and integration of stellar mass functions ($\rho_{\phi}$). A fraction of the total stellar mass is also given for each subclass and method. Quoted error bars are derived according to a jackknife resampling procedure as decribed in \S \ref{mbreakdown}. Derived stellar mass density estimates should also be subject to an additional 22.3\% error contribution from cosmic variance, estimated by the method of \citet{DR2010}.}
\begin{tabular}{ccccc} \hline
Population & $\rho_{\Sigma}/10^7$ & Fraction of All (sum)& $\rho_{\phi}/10^7$ & Fraction of All (fit)\\
           & (M$_{\odot}$Mpc$^{-3}$h$_{0.7}$) &              & (M$_{\odot}$Mpc$^{-3}$h$_{0.7}$) &  \\ \hline
All &$24 \pm 7.9$ & ... & $25 \pm 5.6$ & ... \\
 \hline
E &$8.3 \pm 2.9$ & $0.34$ & $8.7 \pm 4.3$ & $0.35$ \\
S0-Sa &$9.2 \pm 3.0$ & $0.38$ & $8.6 \pm 1.2$ & $0.34$ \\
Sab-Scd &$5.1 \pm 1.6$ & $0.21$ & $5.4 \pm 0.63$ & $0.22$ \\
Sd-Irr &$1.3 \pm 0.40$ & $0.052$ & $1.9 \pm 0.29$ & $0.074$ \\
LBS &$0.23 \pm 0.069$ & $0.0094$ & $0.45 \pm 0.22$ & $0.018$ \\
 \hline
Spheroid Dominated &$17 \pm 5.9$ & $0.73$ & $18 \pm 3.2$ & $0.71$ \\
Disk Dominated &$6.3 \pm 2.0$ & $0.26$ & $6.2 \pm 0.82$ & $0.25$ \\

\hline
\end{tabular}
\end{table*}

\section{Summary and Conclusions}
\label{conc}

We present a significant expansion of the set of galaxy morphological classifications available for a local, volume-limited sample of the GAMA survey, from 3,727 objects in GAMA I to 7,556 objects in the current GAMA-II sample. Using this volume-limited GAMA-II sample ($z < 0.06$ and $r < 19.8$ mag), we derive an updated set of GAMA stellar mass functions, including a breakdown of the total stellar mass function into constituent parts by galaxy morphology. We find broadly consistent results with the GAMA-I stellar mass function analysis of \citet{Kelvin_mfunc}, although some differences exist in the detailed mass function shapes derived for individual galaxy populations.

We find that all individual morphologically defined galaxy classes have mass functions that are well described by a single Schechter function shape and that the total galaxy stellar mass function of our sample is roughly consistent with a double Schechter function shape, although the sum of all individual morphological-type stellar mass functions provides an unsurprisingly closer correspondence to the data. {\color{black}{By extending the mass limit of our sample}} one dex lower than \citet{Kelvin_mfunc}, we are also able to constrain the stellar mass function of the ``little blue spheroid'' (LBS) galaxy population, which was previously not possible due to the primarily low stellar mass nature of this population. This low-mass and blue but apparently bulge-dominated class is intriguing, and it is currently unclear whether they are truly classical spheroid systems or may host disk-like structures that are difficult to observe and/or resolve with relatively shallow and low-resolution SDSS imaging. Deeper and higher-resolution optical imaging from the Kilo-Degree Survey (KiDS; \citealp{kids}), soon to be available for the GAMA survey regions, will allow us to examine this topic further. These data should allow both refinement of current low-redshift classifications and the expansion of GAMA classifications to approximately twice our current upper redshift limit.

We also derive estimates of the total stellar mass density of local galaxies and stellar mass density of individual galaxy classes within our sample. We find a value of $\Omega_{\rm stars} = 0.0018$ relative to the critical density. Taken together our ``dwarf'' Sd-Irr and LBS galaxy classes account for less than 10\% of the total stellar mass density we observe, however it is now well known that such low mass objects can harbor large gas reservoirs and are in fact commonly gas dominated by mass (e.g., \citealp{K04}). Thus, these low-mass objects likely harbour an even more significant fraction of the local \emph{baryon} density, and future large HI surveys such as DINGO will be crucial to illuminating the full census of baryonic mass in local galaxies.

We find approximately 70\% of the local stellar mass density in galaxies that are dominated by a spheroidal component (E and S0-Sa classes), while the remaining approximately 30\% resides in systems with a dominant disk component (Sab-Scd and Sd-Irr classes), consistent with the prior results of \citet{Kelvin_mfunc}. However, if we assume bulge-to-total ratios for our multi-component populations from \citet{GW2008}, we find that roughly half the local stellar mass density is found in each of pure spheroid and disk structures. Taken at face value, this would imply that stellar mass growth up to $z \sim 0$ is roughly equally divided between dissipational and dissipationless processes, at least in the general mixed environment of a volume-limited sample. To explore this topic further, we are also pursuing quantitative bulge-to-disk decompositions of this GAMA-II sample (Lange et al., in prep), allowing us to better quantify the total stellar mass contribution from separable spheroid and disk compononents. Further, it is clear that the overall stellar mass budget will be affected by the mass segregation of galaxies within different large-scale galaxy environments, and in future work we aim to quantify the effects of environment on the growth of stellar mass in spheroidal and disk-like galaxy structures.

\section*{Acknowledgements}

AJM acknowledges funding support from the Australian Research Council. SB acknowledges the funding support from the Australian Research Council through a Future Fellowship (FT140101166).

GAMA is a joint European-Australasian project based around a spectroscopic campaign using the Anglo-Australian Telescope. The GAMA input catalogue is based on data taken from the Sloan Digital Sky Survey and the UKIRT Infrared Deep Sky Survey. Complementary imaging of the GAMA regions is being obtained by a number of independent survey programmes including GALEX MIS, VST KiDS, VISTA VIKING, WISE, Herschel-ATLAS, GMRT and ASKAP providing UV to radio coverage. GAMA is funded by the STFC (UK), the ARC (Australia), the AAO, and the participating institutions. The GAMA website is http://www.gama-survey.org/ .

Funding for the SDSS and SDSS-II has been provided by the Alfred P. Sloan Foundation, the Participating Institutions,
the National Science Foundation, the U.S. Department of Energy, the National Aeronautics and Space
Administration, the Japanese Monbukagakusho, the Max
Planck Society, and the Higher Education Funding Council
for England. The SDSS Web Site is http://www.sdss.org/. The SDSS is managed by the Astrophysical Research
Consortium for the Participating Institutions. The Participating
Institutions are the American Museum of Natural
History, Astrophysical Institute Potsdam, University of
Basel, University of Cambridge, Case Western Reserve University,
University of Chicago, Drexel University, Fermilab,
the Institute for Advanced Study, the Japan Participation
Group, Johns Hopkins University, the Joint Institute for
Nuclear Astrophysics, the Kavli Institute for Particle Astrophysics
and Cosmology, the Korean Scientist Group, the
Chinese Academy of Sciences (LAMOST), Los Alamos National
Laboratory, the Max-Planck-Institute for Astronomy
(MPIA), the Max-Planck-Institute for Astrophysics (MPA),
New Mexico State University, Ohio State University, University
of Pittsburgh, University of Portsmouth, Princeton
University, the United States Naval Observatory, and the
University of Washington.

The VIKING survey is based on observations with ESO
Telescopes at the La Silla Paranal Observatory under the
programme ID 179.A-2004.

\bibliographystyle{mnras}
\bibliography{massfuncs}

%\bsp	% typesetting comment
\label{lastpage}
\end{document}